\documentclass[aps,prl,reprint,superscriptaddress]{revtex4-1}
\usepackage[utf8]{inputenc}

\usepackage{graphicx} 
\usepackage{xcolor}
\usepackage{bm}       

\usepackage{amssymb} 
\usepackage{amsmath}
\usepackage{amsthm}
\usepackage{amsfonts}
\usepackage{mathtools}
\usepackage{ragged2e}
\usepackage{tikz}
\usepackage[hidelinks]{hyperref}
\usepackage[capitalize]{cleveref}

\begin{document}

\title{Better bounds for low-energy product formulas}

\author{Kasra Hejazi} 
\affiliation{Xanadu, Toronto, ON, M5G2C8, Canada}

\author{Modjtaba Shokrian Zini}
\affiliation{Xanadu, Toronto, ON, M5G2C8, Canada}

\author{Juan Miguel Arrazola}
\affiliation{Xanadu, Toronto, ON, M5G2C8, Canada}

\begin{abstract}
Product formulas are one of the main approaches for quantum simulation of the Hamiltonian dynamics of a quantum system. 
Their implementation cost is computed based on error bounds which are often pessimistic, resulting in overestimating the total runtime.
In this work, we rigorously consider the error induced by product formulas when the state undergoing time evolution lies in the low-energy sector with respect to the Hamiltonian of the system. We show that in such a setting, the usual error bounds based on the operator norm of nested commutators can be replaced by those restricted to suitably chosen low-energy subspaces, yielding tighter error bounds.
Furthermore, under some locality and positivity assumptions, we show that the simulation of generic product formulas acting on low-energy states can be done asymptotically more efficiently when compared with previous results. 
\end{abstract}

\maketitle

Simulating the Hamiltonian dynamics of a quantum system is a prominent subroutine in many quantum algorithms, ranging from energy estimation \cite{kitaev1995quantum,cleve1998quantum,lin2022heisenberg} and quantum dynamics \cite{flannigan2022propagation}, to matrix inversion \cite{harrow2009quantum} and quantum adiabatic optimization \cite{albash2018adiabatic}. Several methods for quantum simulation have been developed that include product formulas \cite{lloyd1996universal,suzuki1991general}, truncated Taylor series \cite{berry2015simulating}, and qubitization with quantum signal processing \cite{low2017optimal,low2019hamiltonian}. 

The performance of these algorithms has been studied extensively in general settings, as usually time evolution is the most costly component of the algorithms. For such cost estimation, characteristics of the Hamiltonian of the problem are often taken into account, while the specific properties of the initial state that undergoes time evolution are usually ignored, as illustrated by the very few existing works that actually consider the specifics of the initial state \cite{csahinouglu2021hamiltonian,su2021nearly}. However, it is very likely for these initial states to be highly structured, for example by having support only on low-energy states or having a finite correlation length.

This suggests that in practical settings, a less costly quantum simulation could be feasible if the above-mentioned properties are utilized. As a result, it is an important task to find novel algorithms or estimate the costs of existing ones when more information about the initial state is available. 

In this work, we focus on product formulas as the method of quantum simulation for initial states with negligible support over high-energy eigenstates. Such states constitute an important set of applications, e.g.,~in ground-state energy estimation \cite{kitaev1995quantum,cleve1998quantum,lin2022heisenberg}, computation of low-energy properties of many-body systems \cite{jordan2012quantum,tong2021fast}, and quantum adiabatic algorithms \cite{albash2018adiabatic}. The characterization of the cost for product formulas is often based on the simulation error, and extensive error analysis in generic cases of product formulas has been performed \cite{childs2021theory}. 

The problem of product formulas as applied to low-energy states has also been considered in Ref.~\cite{csahinouglu2021hamiltonian}, where it is shown that under certain assumptions, the first order Trotter-Suzuki formulas can induce a lower cost when applied to low-energy states. 
However, Ref.~\cite{csahinouglu2021hamiltonian} does not observe cost reduction for orders higher than one; in fact, the cost scaling with system size becomes worse when compared with the generic bounds in \cite{childs2021theory}. 

We develop a general framework for studying the error of a product formula for low-energy states. 
We show that we can modify the error bounds of Ref.~\cite{childs2021theory}, which are based on nested commutators of the Hamiltonian terms, so that all the norms are calculated in suitably chosen low-energy subspaces, rather than the whole Hilbert space.
Based on this framework and imposing some further assumptions similar to those in~\cite{csahinouglu2021hamiltonian}, we show that the asymptotic scaling of the cost with system size can in fact be improved for all orders, contrary to the findings in~\cite{csahinouglu2021hamiltonian}. We obtain a better cost scaling due to our tighter error analysis. A summary of asymptotic cost formulas comparing our work, the general cost scaling of Ref.~\cite{childs2021theory}, and previous work on low-energy product formulas~\cite{csahinouglu2021hamiltonian} is shown in Table \ref{tab:comparing_results}.

The rest of the paper is organized as follows. First, we discuss the setting in which our results hold. Then we present our findings on product formula errors for low-energy states, and briefly describe the ideas behind the proofs, with details in the Supplemental Material \cite{SM}. A crucial step is the calculation of the leakage from low to higher energies due to the action of the product formulas, which can be of independent interest. We then use the derived low-energy error bounds to estimate the cost of time evolution for low-energy states. We conclude with a discussion of the results and future directions.

\begin{table*}[]
    \centering
    \begin{tabular}{|l|l|l|l|}
    \hline
         Order & General, Ref.~\cite{childs2021theory} & Low-energy, Ref.~\cite{csahinouglu2021hamiltonian} & Low-energy, present work \\
         \hline
         $1$ & $\mathcal{O} \left(\frac{T^2 N }{ \epsilon } \right)$ & $\tilde{\mathcal{O}}\left( \frac{T^2 \Delta^2 }{ \epsilon} + \frac{T^{4/3} N^{2/3} }{ \epsilon^{1/3} } \right)$ &  $\tilde{\mathcal{O}}\left( \frac{T^2 \Delta^2 }{ \epsilon} + \frac{T^{6/5} N^{2/5} }{ \epsilon^{1/5} } \right)$  \\
         \hline
         $2$ & $\mathcal{O} \left( \frac{T^{3/2} N^{1/2} }{ \epsilon^{1/2} } \right)$ & $\tilde{\mathcal{O}}\left( \frac{T^{3/2} \Delta^{3/2} }{ \epsilon^{1/2}} + \frac{T^{6/5} N^{3/5} }{ \epsilon^{1/5} } \right)$ & $\tilde{\mathcal{O}}\left( \frac{T^{3/2} \Delta^{3/2} }{ \epsilon^{1/2}} + \frac{T^{12/11} N^{3/11} }{ \epsilon^{1/11} } \right)$  \\
         \hline
         $3$ & $\mathcal{O} \left( \frac{ T^{4/3} N^{1/3} }{\epsilon^{1/3} } \right)$ &  $\tilde{\mathcal{O}}\left( \frac{T^{4/3} \Delta^{4/3} }{ \epsilon^{1/3}} + \frac{T^{8/7} N^{4/7} }{ \epsilon^{1/7} } \right)$ &  $\tilde{\mathcal{O}}\left( \frac{T^{4/3} \Delta^{4/3} }{ \epsilon^{1/3}} + \frac{T^{20/19} N^{4/19} }{ \epsilon^{1/19} } \right)$   \\
         \hline
    \end{tabular}
    \caption{Comparing the costs (number of Trotter steps) between the general case bounds of Ref.~\cite{childs2021theory}, the low-energy results from Ref.~\cite{csahinouglu2021hamiltonian}, and the present work. Hamiltonians are taken to consist of positive semidefinite terms for this comparison. The scalings in the low-energy cases are correct up to polylogarithmic factors in the parameters appearing in the arguments of $\tilde{\mathcal{O}}$. The parameters $T, N, \epsilon$, and $\Delta$ denote total evolution time, the system size, the simulation error, and the energy cutoff of the initial state, respectively.}
    \label{tab:comparing_results}
\end{table*}

\section{Setting}

We consider time-independent qubit Hamiltonians written in the form $H=\sum_{m=1}^M H_m$. Each term $H_m$ comprises a number $L_m$ of $k$-local interactions that commute with each other; while $L_m$ can scale with the total number of qubits $N$, we take $M$ to be $O(1)$ and let $L = \max_m L_m$.
Each qubit is involved in at most $d$ of the $k$-local interactions. Furthermore, 
we introduce the parameter $J$ such that the norm of each local interaction is upper bounded by $J/2$.
\footnote{More precisely, $J := 2\max_h \min_c \| h+c \|$, with $h$ running over single local terms of the Hamiltonian and $c$ a constant, see \cite{SM}}. 

We would like to bound the error of a product formula written in the form
\begin{equation}\label{eq:general_product_formula}
    \mathcal{W}(s) = e^{-i s_q H_{m_q}} \ldots e^{-i s_2 H_{m_2}} \, e^{-i s_1 H_{m_1}},
\end{equation}
that approximates time evolution $e^{-iHs}$ under $H$. Here, $m_1,\ldots,m_q$ is a permutation of $1,\ldots,M$, with repetitions in the case of higher-order product formulas, with `order' defined further below. Each $s_j = a_j s$ is proportional to $s$ with $a_j \in \mathbb{R}$ and the total number $q$ of fast-forwardable exponentials in Eq.~\eqref{eq:general_product_formula} is taken to be $O(1)$.

The operator $\mathcal{W}(s)$ in Eq.~\eqref{eq:general_product_formula} implements the time evolution for a single Trotter step of length $s$; evolving for total time $T$ can be performed through a number $r=T/s$ of Trotter steps. If the error in a single step is $\varepsilon(s)$, the total accumulated error will be $r \varepsilon(s)$. We characterize the cost in terms of the Trotter number $r$; therefore, a tighter error estimation is a proxy to a lower cost. 

The order of a product formula is defined based on the error that is accumulated in a single Trotter step: a product formula of order $p$ has error scaling $\varepsilon(s) = \mathcal{O}(s^{p+1})$. More precisely, the operator error satisfies \cite{childs2021theory}:
\begin{equation}\label{eq:error_operator}
\begin{aligned}
    \left\| e^{-iHs}- \mathcal{W}(s) \right\| = \mathcal{O} \big( s^{p+1} \, \sum_{m=1}^M \sum_{\{m_i\} } f_{\{m_i\},m}  \\
    \times   \left\|  [H_{m_p}, \ldots ,[H_{m_1},H_m]]\ldots]  \right\| \big),
\end{aligned}    
\end{equation}
where the $m$ and $\{m_i\}={m_1,\ldots,m_p}$ summations run over all terms of the Hamiltonian. 
The coefficients $f_{\{m_i\},m}$ are $\mathcal{O}(1)$ positive numbers that are determined based on the product formula, i.e.,~the values of $a_j$.

We present our general results based on the above settings and then specialize to the case in which we take all individual terms $H_m$ to be positive semidefinite. Note that with the latter assumption, the setting becomes similar to that of Ref.~\cite{csahinouglu2021hamiltonian}. 
Furthermore, this is the scenario for which the results of Table \ref{tab:comparing_results} are computed using our method.
As discussed in Ref.~\cite{csahinouglu2021hamiltonian}, the assumption of positive semidefinite holds for certain classes of spin Hamiltonians, e.g. frustration-free systems \cite{bravyi2010complexity}.

As we will discuss later, the setting can also be extended to include fermionic systems in which the Hamiltonian consists of a sum of k-local terms made of an even number of creation and annihilation fermionic operators, with each fermionic mode participating in at most $d$ interactions. Notably, this extends the results of this paper to wide classes of fermionic systems such as electronic structure Hamiltonians.

\section{Low-energy Error}
We calculate the error for the product formula in Eq.~\eqref{eq:general_product_formula} acting on a state with negligible support beyond an energy cutoff $\Delta$. We characterize this error by calculating the following operator norm for a single Trotter step of length $s$. All norms are spectral norms in this work.
\begin{equation}\label{eq:error_low_energy_def}
\begin{aligned}
    \varepsilon_\Delta(s) &= \left\| \left( \mathcal{W}(s) - e^{-iHs} \right) \Pi_{\leq \Delta} \right\|.
\end{aligned}
\end{equation}
We use the notation $\Pi_{\leq \Delta}$ and $\Pi_{>\Delta'}$ to denote the projectors onto the subspaces spanned by eigenstates of $H$ with energy $E \leq \Delta$ and $E>\Delta'$. We occasionally abuse this notation to also refer to the subspaces themselves. We also use the notation $\| A \|_{\leq \Lambda}$ to represent the restricted norm $\| \Pi_{\leq \Lambda} \; A \; \Pi_{\leq \Lambda} \|$ to such low-energy subspaces.

One of our main results for bounding $\varepsilon_\Delta$ is the following:  
the right-hand side of Eq.~\eqref{eq:error_operator} maintains its accuracy in error representation for initial states with negligible support on $\Pi_{>\Delta}$, if all the norms are replaced by restricted norms to a suitably chosen low-energy subspace; this amounts to using norms of {\it projected} nested commutators $\|  [H_{m_p}, \ldots ,[H_{m_1},H_m]]  \|_{\leq\Delta'}$ for some appropriate value of $\Delta'$, in the right-hand side of Eq.~\eqref{eq:error_operator}.
This modification will result in a lower overall error and estimated cost as projected nested commutators have smaller norms.
We identify suitable $\Delta'$ values in the following.

\textbf{Separating the error terms.} We now turn to the study of $\varepsilon_\Delta(s)$. Using the triangle inequality, this error is upper bounded by:
\begin{equation}\label{eq:breaking_the_error}
\begin{aligned}
    \left\| \Pi_{\leq \Delta'} \left( \mathcal{W}(s) - e^{-iHs} \right) \Pi_{\leq \Delta} \right\| 
    + \left\| \Pi_{> \Delta'}  \mathcal{W}(s)  \Pi_{\leq \Delta} \right\|,
\end{aligned}
\end{equation}
with $\Delta'>\Delta$ to be determined later.
We dropped the exact time evolution operator $e^{-iHs}$ in the second term, as it commutes with the two projectors and $\Pi_{> \Delta'}\Pi_{\leq \Delta}=0$.

The above equation implies that the error can be calculated as a sum of two separate terms: a part confined to energies below $\Delta'$, which we dub the \textit{retained} component, and the leakage by $\mathcal{W}(s)$ to $\Pi_{>\Delta'}$. Below, we first state the final bounds for these two terms for an order $p$ product formula, and then discuss the ideas behind their proofs. 

\textbf{Product formula leakage.} We show in \cite{SM} that:
\begin{equation}\label{eq:leakage_product_final_form}
    \left\| \Pi_{>\Delta'}  \mathcal{W}(s) \Pi_{\leq \Delta} \right\| \leq  e^{-\lambda \left[\Delta' - \Delta - 2 R^{\mathcal{W}}(s) \right]}   ,
\end{equation}
with $\lambda = \frac{1}{2Jdk}$ and $R^{\mathcal{W}}(s)$ a positive number, which for an order $p$ product formula takes the form (see \cite{SM} for details):
\begin{equation}\label{eq:RI_to_lowest_order}
    R^{\mathcal{W}}(s) = \alpha \, s^{p+1} L + \mathcal{O}(L s^{p+2}),
\end{equation}
with $\alpha=  \frac{\ell_p}{\lambda(p+1)} \sum_{m,\{m_i\}} |f_{\{m_i\},m}|$ and $\ell_p = L \, p! \left( Jkd e \right)^p eJ$.
We note here that this scaling is better than the scaling found in \cite{csahinouglu2021hamiltonian} which is $R^{\mathcal{W}} = \mathcal{O}(Ls)$, as $s$ is considered small.

\textbf{The retained component.} The first term in Eq.~\eqref{eq:breaking_the_error} can be generally bounded as follows, as we show in \cite{SM}:
\begin{equation}\label{eq:error_operator_retained_proj}
\begin{aligned}
    \left\| \Pi_{\leq \Delta'} \left( \mathcal{W}(s) - e^{-iHs} \right) \Pi_{\leq \Delta} \right\|  = \qquad \qquad \qquad \qquad & \\
    \mathcal{O}  \left( s^{p+1} \, \sum_{m,\{ m_i\}}  \left\| [H_{m_p}, \ldots ,[H_{m_1},H_m]]\ldots]  \right\|_{\leq \Delta'}
     \right), 
\end{aligned}    
\end{equation} 
with $\Delta'$ chosen large enough so that $\Delta' - \Delta = \mathcal{O}(s^{p+1} L)$ up to polylogarithmic corrections in $L$ and $s$. 

From the above two bounds, the total error $\varepsilon_\Delta(s)$ can also be bounded as follows: we note that the leakage component shown in Eq.~\eqref{eq:leakage_product_final_form} can also become subdominant to Eq.~\eqref{eq:error_operator_retained_proj}, provided that $\Delta'-\Delta = \mathcal{O}(s^{p+1} L)$ up to polylogarithmic corrections in $L$ and $s$; interestingly this is the same requirement for Eq.~\eqref{eq:error_operator_retained_proj} to hold. This means that with such $\Delta'$, the whole low-energy error $\varepsilon_\Delta(s)$ of Eq.~\eqref{eq:error_low_energy_def} is dominated by the right-hand side of Eq.~\eqref{eq:error_operator_retained_proj}.

Remarkably, this is showing that $\varepsilon_\Delta$ can be calculated with the use of operator norms of {\it restricted} nested commutators into a subspace of energies of at most $\Delta'$, provided $\Delta'$ is chosen according to the above discussion. Note that this projection results in tighter bounds when contrasted with taking the operator norm without projection as in the general result of Eq.~\eqref{eq:error_operator}. We use this error analysis below for the case of Hamiltonians with positive semidefinite terms and calculate the resulting implementation cost later in the text.

\textbf{Positive semidefinite Hamiltonian terms.} The retained component bound in Eq.~\eqref{eq:error_operator_retained_proj} can be made more concrete when the $H_m$'s are positive semidefinite \cite{SM}:
\begin{equation}\label{eq:error_retained component}
    \left\| \Pi_{\leq \Delta'} \left( \mathcal{W}(s) - e^{-iHs} \right) \Pi_{\leq \Delta} \right\|  = \mathcal{O}(s^{p+1} \Delta_f^{p+1}),
\end{equation}
where $\Delta_f>\Delta'$ is also introduced to suppress more leakage norms, and it is required that both $\Delta'$ and $\Delta_f$ equal $\Delta +  \mathcal{O}(s^{p+1} L)$ up to polylogarithmic corrections in $L$ and $s$ ($\Delta'$ needs to be larger than $\Delta_f$ only by polylogarithmic factors in $L$ and $s$.).

Again, the leakage error in Eq.~\eqref{eq:leakage_product_final_form} becomes subdominant with respect to the error in Eq.~\eqref{eq:error_retained component} with the same condition on $\Delta'$.
This bounds the error in each step as:
\begin{equation}\label{eq:error_single_step}
    \varepsilon_\Delta(s) = \mathcal{O}(s^{p+1} \Delta_f^{p+1}),
\end{equation}
with $\Delta_f - \Delta = \mathcal{O}(s^{p+1}  L)$ up to polylogarithmic factors in $L$ and $s$.

\section{Ideas behind the proofs}
Let us now discuss the ideas behind the proofs. First, similar to \cite{csahinouglu2021hamiltonian}, we prominently use theorem 2.1 of \cite{arad2016connecting}, which allows one to bound the leakage by some arbitrary $A$ from energies below some cutoff $\Lambda$ to those higher than some other cutoff $\Lambda'>\Lambda$:
\begin{equation}\label{eq:leakage_generic_A}
    \left\| \Pi_{>\Lambda'} \, A \, \Pi_{\leq \Lambda}\right\| \leq \| A \| \ e^{-\lambda (\Lambda'-\Lambda - 2R) },
\end{equation}
where $\lambda = \frac{1}{2gk}$,  $g$ an upper bound for the sums of strengths of all terms in the Hamiltonian involving a qubit, and  $R$ is the sum of the norms of all local terms in the Hamiltonian that do not commute with $A$. More concretely, given the types of Hamiltonians assumed in this work, $g=Jd$, and for a single $k$-local term in the Hamiltonian, $R$ can be upper bounded by $kJd$.

Ref.~\cite{csahinouglu2021hamiltonian} also uses the same leakage bound for estimating $\varepsilon_\Delta(s)$ in Eq.~\eqref{eq:error_low_energy_def}: the leakage due to the product formula is calculated through bounding the leakage of each single exponential in the expression Eq.~\eqref{eq:general_product_formula} and then summing all of the contributions. We show that this overestimates the leakage and hence the total error. Concretely, the leakage happens according to Eq.~\eqref{eq:leakage_product_final_form} with $R^{\mathcal{W}} = \mathcal{O}(s^{p+1} L)$ instead of the scaling  $R^{\mathcal{W}} = \mathcal{O}( s L)$ of \cite{csahinouglu2021hamiltonian}; this directly results in the better cost scalings shown in Table \ref{tab:comparing_results}.
To reach this result, we take a different approach for bounding the leakage by noting that the product of all the exponentials in Eq.~\eqref{eq:general_product_formula} is close to the exact time evolution operator $e^{-iHs}$, which does not have any leakage; in fact, the two only differ by terms that scale as $\mathcal{O}(s^{p+1})$ for an order $p$ product formula \cite{childs2021theory}. 
Therefore, we focus on bounding directly the leakage of the entire product instead of each single exponential, and subsequently obtain tighter bounds for this error.

To implement this approach, we use the theory of Trotter error developed in \cite{childs2021theory} and write $\mathcal{W}(s)$ as a time-ordered integral:
\begin{equation}\label{eq:time_ordered_w}
    \mathcal{W}(s) = \mathcal{T} e^{-i \int_0^s  d\sigma \;\left( H +  \mathcal{E}(\sigma)  \right) },
\end{equation}
where $\mathcal{T}$ is the time ordering operator and $\mathcal{E}(\sigma)$ is written solely in terms of nested commutators of the terms of the Hamiltonian.
As we show in \cite{SM}, the product formula error $\|e^{-iHs} - \mathcal{W}(s)\|$ can be upper bounded by $\int_0^s d\sigma \; \left\|\mathcal{E}(\sigma) \right\|$, which results in the same nested commutator error bounds of \cite{childs2021theory}.
This is because for a product formula of order $p$, the nested commutators appearing in $\mathcal{E}(\sigma)$ are of order at least $p$ and $\mathcal{E}(\sigma) = \mathcal{O}(\sigma^p)$; indeed, to lowest order in $s$, $\int_0^s d\sigma \; \left\|\mathcal{E}(\sigma) \right\|$ results in Eq.~\eqref{eq:error_operator} (details in \cite{SM}).

Using Eq.~\eqref{eq:time_ordered_w}, expanding the time ordered exponential as a Dyson series and analyzing terms separately, we show that the leakage due to the product formula is given by Eq.~\eqref{eq:leakage_product_final_form}. In particular, since all the terms in $\mathcal{E}$ are nested commutators of the terms of the Hamiltonian, and thus all of them scale linearly with system size, we observe that $R^{\mathcal{W}}(s)$ in Eq.~\eqref{eq:RI_to_lowest_order} also scales linearly with system size. Indeed, the fact that the lowest order contribution to $R^{\mathcal{W}}(s)$ is $\mathcal{O}(s^{p+1})$ in Eq.~\eqref{eq:RI_to_lowest_order} is also a direct consequence of $\mathcal{E}(\sigma)$ being $\mathcal{O}(\sigma^p)$.

For the retained component, a similar approach applies, where we show in \cite{SM} that $\left\| \Pi_{\leq \Delta'} \left( \mathcal{W}(s) - e^{-iHs} \right) \Pi_{\leq \Delta} \right\|$ can be upper bounded by:
\begin{equation}\label{eq:int_projected_E}
    \int_0^s d\sigma \; \left\| \mathcal{E}(\sigma)   \right\|_{\leq\Delta'} + \xi_1 \left\| \Pi_{>\Delta'}  \mathcal{W}(s) \Pi_{\leq \Delta} \right\|.
\end{equation}
with $\xi_1 = \mathcal{O}(s^{p+1} L)$.
Interestingly, with taking $\Delta'-\Delta = \mathcal{O}(s^{p+1}L)$ up to polylogarithmic factors, the second term, which is similar to the above leakage term, becomes subdominant with respect to $\int_0^s d\sigma \; \left\| \mathcal{E}(\sigma)   \right\|_{\leq\Delta'}$, reducing the sum to the right-hand side of Eq.~\eqref{eq:error_operator_retained_proj} to lowest order is $s$. Note that the same condition on $\Delta'$ also suppresses the leakage term in Eq.~\eqref{eq:breaking_the_error}, thus making $\int_0^s d\sigma \; \left\| \mathcal{E}(\sigma)   \right\|_{\leq\Delta'}$ dominate the error $\varepsilon_{\Delta}$, which is as stated before one of our main results. This is opposed to $\int_0^s d\sigma \; \left\|\mathcal{E}(\sigma) \right\|$, which was mentioned above as the whole operator error.

Furthermore, when terms $H_m$ of the Hamiltonian are positive semidefinite, Eq.~\eqref{eq:error_operator_retained_proj} can result in Eq.~\eqref{eq:error_retained component} 
by resolving the nested commutator (i.e.~using $[X,Y] \to XY - YX$) step by step; this is done by breaking every resolved commutator into a leakage term and a retained component by including a series of cutoffs $\Delta' < \Delta_1 <\ldots<\Delta_p = \Delta_f$, and choosing them in a way so that all the leakages are negligible. 
As we show in \cite{SM}, if $\Delta_f - \Delta = \mathcal{O}(s^{p+1} L)$ up to polylogarithmic corrections, all the leakage terms will be suppressed. 
With all leakage terms negligible, the norm of every order $p$ nested commutator essentially reduces to a product of a number $p$ of {\it restricted} norms such as $\| H_m \|_{\leq \Delta_f}$; each one of these can be bounded by $\Delta_f$ because of their positive semi-definiteness. A similar discussion for non-positive-semidefinite $H_m$ terms is presented in \cite{SM}.

One last point to stress here is that all the steps of the proofs also hold for fermionic Hamiltonians whose single terms consist of at most $k$ fermionic operators and each fermionic mode enters at most $d$ interactions. Crucially, the main theorem of \cite{arad2016connecting}, i.e.,~Eq.~\eqref{eq:leakage_generic_A}, also holds for such fermionic Hamiltonians, as the entire proof only depends on the commutativity of Hamiltonian terms and the operator $A$. This generalizes the above results to fermionic Hamiltonians as well.

\section{The cost}
\begin{figure}[!t]
    \centering
    \includegraphics[width = 0.4\textwidth]{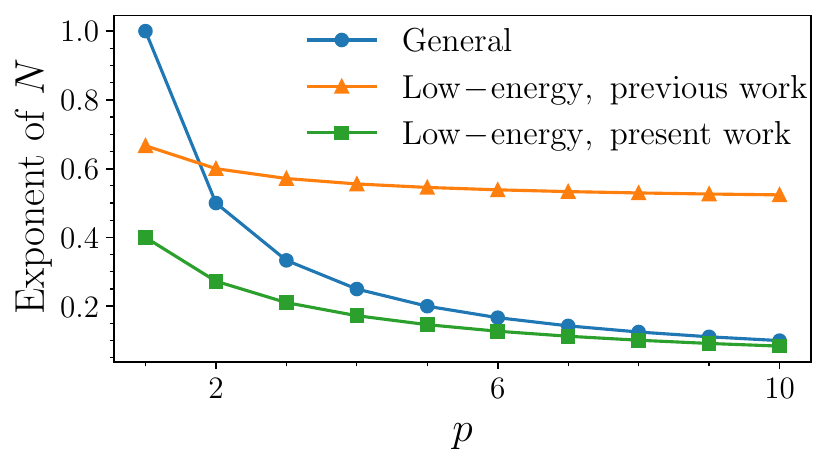}
    \caption{Scaling of the exponent of the system size $N$ in the cost (Trotter number) based on the general case \cite{childs2021theory}, previous low-energy results \cite{csahinouglu2021hamiltonian}, and the present work. The best scaling corresponds to the present work.}
    \label{fig:n_exponents}
\end{figure}
We focus on Hamiltonians with positive semidefinite terms here. We discussed above how the error of a single step can be bounded as in Eq.~\eqref{eq:error_single_step}, with $\Delta_f - \Delta = \mathcal{O}(s^{p+1}N)$  up to polylogarithmic factors in $s$ and $N$, for an order $p$ product formula; $L$ is assumed to scale as the system size $N$. 
For a quantum simulation of time $T$, with total tolerated error $\epsilon$, we break the time evolution into $r$ segments, each with length $s=T/r$ and require that the sum of the errors of segments is equal to $\epsilon$, i.e.~$\varepsilon_\Delta \left(T/r\right) r = \epsilon$.

In order to find asymptotic cost scaling, we take the effect of the two terms in $\Delta_f$ separately (assuming each one is the dominant source of cost in turn), and find an asymptotic form for the required number of Trotter steps $r$ so that the error stays at the tolerated level. Using Eq.~\eqref{eq:error_single_step},
we consider a Trotter number $r$ that satisfies both $\left[(T/r) \, \Delta\right]^{p+1} = \mathcal{O}(\epsilon/r), \ \left[(T/r) \, (\Delta_f-\Delta) \right]^{p+1} = \mathcal{O}(\epsilon/r)$, which results in:
\begin{equation}\label{eq:cost_low_energy_present}
    r = \mathcal{O} \left( 
    \frac{(T \Delta)^{1+1/p}}{\epsilon^{1/p}}
    + T \left[\frac{T  N^{p+1} }{\epsilon}\right]^{\frac{1}{(p+1)^2+p}}
    \right),
\end{equation}
up to polylogarithmic factors in the parameters.

Now, we assume that $\Delta$ has a slow enough growth with $N$, such that the main scaling of the cost with $N$ originates from the second term in Eq.~\eqref{eq:cost_low_energy_present}.
This scaling, which reads $N^{1/(p+1+\frac{p}{p+1})}$, is better than the general cost scaling of $N^{1/p}$ \cite{childs2021theory}, and the previous low-energy result $N^{1/2+1/(4p+2)}$ \cite{csahinouglu2021hamiltonian}. Note that the previous low-energy result of Ref.~\cite{csahinouglu2021hamiltonian} only beats the general case in terms of scaling with $N$ for the first order product formulas, while our work provides a better cost for all orders. The advantage is due to the better leakage bound found in Eq.~\eqref{eq:leakage_product_final_form}, where $R^{\mathcal{W}} = \mathcal{O} (s^{p+1} N) $ rather the $\mathcal{O}(sN)$ scaling of \cite{csahinouglu2021hamiltonian}.

A comparison between present results and those in Refs.~\cite{childs2021theory} and \cite{csahinouglu2021hamiltonian} is presented in Table~\ref{tab:comparing_results}. Exponents with respect to $N$ for all the cases are depicted in Fig.~\ref{fig:n_exponents}.

\section{Discussion}
In this work, we establish state-of-the-art error bounds for product formulas acting on low-energy states. In particular, we showed how the previous general nested commutator error bounds of \cite{childs2021theory} can be modified to obtain tighter bounds for low-energy states, i.e.,~by calculating operator norms restricted to lower-energy subspaces.
Furthermore, we showed that for Hamiltonians with positive semidefinite terms, the scaling of the Trotter number with respect to the system size is asymptotically better than the general scaling results \cite{childs2021theory} and previous results in the same setting \cite{csahinouglu2021hamiltonian}. 
More precisely, the previous results of Ref.~\cite{csahinouglu2021hamiltonian} only showed improvement in the cost for the first order Trotter-Suzuki formulas, while the current work shows that the improvement can be obtained for all orders;
this is especially important as second-order Trotter-Suzuki formulas can be utilized with the same resource requirement as the first order.

Even though the error analysis for Hamiltonians with positive semidefinite terms has been considered here, cost estimation based on the more general error analysis in Eq.~\eqref{eq:error_operator_retained_proj} remains an open problem, and we leave it to future work to explore implications of this more general result. It would also be interesting to consider other possible structures in the low-energy states of interest, e.g.,~a finite correlation length \cite{hastings2004lieb,anshu2016concentration} and similar properties. Application of the current bound to time-dependent Hamiltonian simulation problems as those arising in, e.g.,~adiabatic quantum computing, can be an interesting topic to consider.

\bibliography{low.bib}

\newpage

\onecolumngrid

    \centerline{ {\huge {\bf Supplemental material}} }

\bigskip

\section{Setting}

We consider time-independent qubit Hamiltonians written in the form $H=\sum_{m=1}^M H_m$. Each term $H_m$ comprises a number $L_m$ of $k$-local interactions that commute with each other; while $L_m$ can scale with the total number of qubits $N$, we take $M$ to be $O(1)$ and let $L = \max_m L_m$.
Each qubit is involved in at most $d$ of the $k$-local interactions. Furthermore, 
we introduce the parameter $J$ to denote the interaction strength.
More precisely, $J := 2\max_h \min_c \| h+c \|$, with $h$ running over single local terms of the Hamiltonian and $c$ a constant. 

Considering the general product formula:
\begin{equation}\label{eq:sm_general_product_formula}
    \mathcal{W}(s) = e^{-i s_q H_{m_q}} \ldots e^{-i s_2 H_{m_2}} \, e^{-i s_1 H_{m_1}},
\end{equation}
that approximates time evolution $e^{-iHs}$ under $H$. Here, $m_1,\ldots,m_q$ is a permutation of $1,\ldots,M$, with repetitions in the case of higher-order product formulas. Each $s_j = a_j s$ is proportional to $s$ with $a_j \in \mathbb{R}$ and the total number $q$ of simple exponentials in Eq.~\eqref{eq:sm_general_product_formula} is taken to be $O(1)$.

\section{The error operator}
In this section, we discuss general forms for the error operator of a product formula, the discussion is based on Ref.~\cite{childs2021theory}, and we include it here for completeness.

Considering the general product formula in \eqref{eq:sm_general_product_formula},
we can write its derivative obtained by directly differentiating \cite{childs2021theory} as:
\begin{equation}\label{eq:sm_deriv_W}
    \frac{d}{ds} \mathcal{W}(s) = -i \mathcal{F}(s) \, \mathcal{W}(s),
\end{equation}
where
\begin{equation}
\begin{aligned}
    \mathcal{F}(s) = \sum_{\nu=1}^q a_\nu \ V_\nu H_{m_\nu} V_{\nu}^\dagger, \qquad \qquad 
    V_{\nu} = e^{-i s_q H_{m_q}} \ldots e^{-i s_{\nu+1}H_{m_\nu+1}}.
\end{aligned}    
\end{equation}
Note that we are using $\nu$ to enumerate terms in the permutation of Hamiltonian terms used in Eq.~\eqref{eq:sm_general_product_formula}, while $m_i$ is used to enumerate terms of the Hamiltonian.
Now, $\mathcal{W}$ can be written as a time ordered integral over $\mathcal{F}$:
\begin{equation}\label{eq:sm_time_ordered_w}
    \mathcal{W}(s) = \mathcal{T} e^{-i \int_0^s  d\sigma \; \mathcal{F}(\sigma) }.
\end{equation}

Using the Hadamard formula:
\begin{equation}\label{eq:hadamard_formula}
\begin{aligned}
    e^{-is A} \, B \, e^{is A} = B + (-is) [A,B] + \frac{(-is)^2}{2!} [A,[A,B]] + \ldots,
\end{aligned}
\end{equation}
we can see that $\mathcal{F}$ is an infinite collection of nested commutators; each order of $s^n$ in $\mathcal{F}$ contains nested commutators of order $n$:
\begin{equation}\label{eq:expansion_of_F_nested}
\begin{aligned}
    \mathcal{F}(s) &= H \\
    &+ (-is) \sum_{\nu=1}^q \sum_{\nu_1=\nu+1}^q \tilde{f}_{\nu_1,\nu} [H_{m_{\nu_1}},H_{m_\nu}]\\
    &+ (-is)^2 \sum_{\nu=1}^q \sum_{\nu_1=\nu+1}^q \sum_{\nu_2=\nu_1}^q \tilde{f}_{\nu_2\nu_1,\nu} [H_{m_{\nu_2}},[H_{m_{\nu_1}},H_{m_\nu}]]\\
    &+ (-is)^3 \sum_{\nu=1}^q \sum_{\nu_1=\nu+1}^q \sum_{\nu_2=\nu_1}^q \sum_{\nu_3=\nu_2}^q \tilde{f}_{\nu_3\nu_2\nu_1,\nu} [H_{m_{\nu_3}},[H_{m_{\nu_2}},[H_{m_{\nu_1}},H_{m_\nu}]]]\\
    &+\ldots .
\end{aligned}
\end{equation}
The coefficients can be calculated as:
\begin{equation}
\begin{aligned}
    \tilde{f}_{\nu_1,\nu} = a_\nu a_{\nu_1}; \qquad &\nu_1> \nu, \\
    \tilde{f}_{\nu_2\nu_1,\nu} = a_\nu a_{\nu_1} a_{\nu_2}; \qquad &\nu_2 > \nu_1> \nu, \qquad \tilde{f}_{\nu_1\nu_1,\nu} = \frac{1}{2!}  a_\nu a_{\nu_1}^2,\\
    \tilde{f}_{\nu_3\nu_2\nu_1,\nu} = a_\nu a_{\nu_1} a_{\nu_2} a_{\nu_3}; \qquad &\nu_3 > \nu_2 > \nu_1> \nu, \qquad \tilde{f}_{\nu_1\nu_1\nu_1,\nu} = \frac{1}{3!}a_\nu a_{\nu_1}^3, \qquad \ldots\\
    \vdots
\end{aligned}
\end{equation}
Note that in all of the above equations, the summation is performed with the indices $\nu$ and $\nu_i$ enumerating the permutations of the Hamiltonian terms occurring in the product formula \eqref{eq:sm_general_product_formula}. However, the summation can also be done over the Hamiltonian term indices $m$ resulting:
\begin{equation}\label{eq:expansion_of_F_nested_m_index}
\begin{aligned}
    \mathcal{F}(s) &= H 
    + (-is) \sum_{m=1}^M \sum_{m_1 \neq m} f_{m_1,m} [H_{m_1},H_m]\\
    &+ (-is)^2 \sum_{m=1}^M \sum_{m_1,m_2\neq m} f_{m_2m_1,m} [H_{m_2},[H_{m_1},H_m]]+\ldots .
\end{aligned}
\end{equation}
with $f$ and $\tilde{f}$ coefficients related through the particular permutation used in Eq.~\eqref{eq:sm_general_product_formula}.

The lowest nonvanishing order of $s$ in the expansion of $\mathcal{F}-H$ controls the error of the product formula \cite{childs2021theory}. For a Trotter Suzuki product formula of order $p$, it is well-known that the error in a single step should scale as $s^{p+1}$; this means that all the terms with nested commutators of orders less than $p+1$ will vanish in the expansion of $\mathcal{F}-H$ \cite{childs2021theory}. 
Indeed, the error induced by product formula can be written as:
\begin{equation}\label{eq:sm_error_operator_norm}
    \left\|e^{-iHs}- \mathcal{W}(s) \right\| \leq \frac{(s)^{p+1}}{p+1} \, \sum_{m=1}^M \sum_{\{m_i\} } f_{\{m_i\},m} \ \left\| [H_{m_p}, \ldots ,[H_{m_1},H_m]]\ldots] \ \right\| + \mathcal{O}(s^{p+2}).
\end{equation}

\section{Leakage of the product formula}\label{sec:leak_product_form}
In order to calculate the leakage due to $\mathcal{W}$ when starting in a low-energy state, we need to switch to the interaction picture of Eq.~\eqref{eq:sm_time_ordered_w} (see Appendix C of Ref.~\cite{childs2021theory}). First we define $\mathcal{F}(s) = H + \mathcal{E}(s)$, in the interaction picture, Eq.~\eqref{eq:sm_time_ordered_w} becomes:
\begin{equation}\label{eq:time_ordered_interaction_picture}
\mathcal{W}(s) = e^{-iHs} \ \mathcal{T}  e^{-i \int_0^s  d\sigma \; \mathcal{E}^{I}(\sigma) },
\end{equation}
where $\mathcal{E}^I(s) = e^{iHs} \, \mathcal{E}(s) \, e^{-iHs}$, which can again be expanded using the Hadamard formula Eq.~\eqref{eq:hadamard_formula} to get a form for $\mathcal{E}^I$ in terms of nested commutators of Hamiltonian terms. In fact an expansion similar to Eq.~\eqref{eq:expansion_of_F_nested_m_index} can also be written for $\mathcal{E}^I$:
\begin{equation}\label{eq:interaction_picture_E_def}
\begin{aligned}
    \mathcal{E}^I(s) &= (-is) \sum_{m, m_1} f^I_{m_1,m} \, [H_{m_1},H_m]\\
    &+(-is)^2 \sum_{m, m_1, m_2} f^I_{m_2m_1,m} \, [H_{m_2},[H_{m_1},H_m]]\\
    &+ \ldots.
\end{aligned}
\end{equation}
Each nested commutator in the above expansion is obtained either directly from Eq.~\eqref{eq:expansion_of_F_nested} or is obtained through a number of commutations of $H$ with lower order commutators in Eq.~\eqref{eq:expansion_of_F_nested}. Thus each coefficient $f^I$ can be written in terms of coefficients $f$ of the same and lower order. 
Crucially, the lowest nonzero order in the expansions of $\mathcal{E} = \mathcal{F} - H$ and $\mathcal{E}^I$ match. As a result, for a Trotter-Suzuki product formula of order $p$, the lowest order nested commutator in Eq.~\eqref{eq:interaction_picture_E_def} is of order $p$.

Note that a Dyson series expansion can be used to write Eq.~\eqref{eq:time_ordered_interaction_picture} as:
\begin{equation}\label{eq:dyson_series}
\begin{aligned}
    \mathcal{W}(s) = e^{-iHs} \Bigg[ 1 + \sum_j \frac{(-i)^j}{j!} 
    \int_0^s d\sigma_1 \int_0^s d\sigma_2 \ldots \int_0^s d\sigma_j \; \mathcal{T} \mathcal{E}^I(\sigma_1)\mathcal{E}^I(\sigma_2) \ldots \mathcal{E}^I(\sigma_j) \Bigg].
\end{aligned}
\end{equation}

Now, we need to turn our attention to leakage from low energies; like Ref.~\cite{csahinouglu2021hamiltonian}, the main result we will be using is Theorem 2.1 of Ref.~\cite{arad2016connecting}. We first explain it here using the present notation: take $\Pi_{ > \Lambda'}$ and $\Pi_{ \leq \Lambda}$ to be projectors onto subspaces spanned by eigenstates of $H$ with energies satisfying $E> \Lambda'$ and $E \leq \Lambda'$ respectively. For an operator A, the following inequality holds for its leakage from a low-energy subspace:
\begin{equation}\label{eq:sm_leakage_generic_A}
    \left\| \Pi_{>\Lambda'} \, A \, \Pi_{\leq \Lambda}\right\| \leq \| A \| \ e^{-\lambda (\Lambda'-\Lambda - 2R) },
\end{equation}
where $\lambda = \frac{1}{2gk}$ with $g$ an upper bound for the sums of strengths of all terms involving each spin, and $R$ is the sum of the norms of all local terms in the Hamiltonian that do not commute with $A$. Given the Hamiltonian assumed in this work (sum of at most $k$-local terms and with maximum degree $d$), $g=Jd$. For a single $k$-local term in the Hamiltonian, $R$ can be upper bounded by $kJd$. \\
\noindent  Note: Ref.~\cite{arad2016connecting} assumes positive semidefinite single terms for the Hamiltonian; this is not a crucial assumption for their theorem to work, 
and in fact their leakage bound proof also hold for nonpositive single terms in the Hamiltonian.
This just means that $J$ should be chosen with care: in the worst case for a nonpositive operator we take $J$ to be twice as large as the norm of the largest single term of the Hamiltonian; for example, if single terms of the Hamiltonian are Pauli strings, $J$ should be taken as twice the value of the largest coefficient. In general $J$ is chosen as $2\max_h \min_c \| h+c \|$, with $h$ running over single local terms of the Hamiltonian and $c$ showing a constant.

Now consider a nested commutator of order $n$ such as $[H_{m_n}, \ldots ,[H_{m_1},H_{m}]]\ldots]$, we would like to calculate its leakage from low energies similar to Eq.~\eqref{eq:sm_leakage_generic_A}. We first note that altogether such a nested commutator consists of at most $L (kd)^n n!$ nested commutators of $k-$local terms (see next section and table \ref{tab:nested_commutator_terms}). Each term within a nested commutator of $k-$local terms has a support of at most $(n+1)k$, strength of at most $ J^{n+1}$ (see next Section and Table \ref{tab:nested_commutator_terms}).
As a result of the previous observation on the length of single terms, the value of $R$ for these nested commutator is upper bounded by $J (n+1) k d$. We can now use Eq.~\eqref{eq:sm_leakage_generic_A} for each single term in this nested commutator to bound the leakage of the whole commutator as follows by summing all of the contributions together:
\begin{equation}\label{eq:leakage_single_nested_commutator}
\begin{aligned}
    \left\| \Pi_{>\Lambda'} \ [H_{m_n}, \ldots ,[H_{m_1},H_m]]\ldots] \ \Pi_{\leq \Lambda}\right\| 
    \leq \ell_n \ e^{-\lambda(\Lambda' - \Lambda)},
\end{aligned}
\end{equation}
where
\begin{equation}\label{eq:defintion_of_ell_n}
\begin{aligned}
    \ell_n &= (L (kd)^n n!) \ ( J^{n+1}) \ e^{2\lambda Jkd(n+1)}\\
    &= L \, n! \left( Jkd e \right)^n eJ.
\end{aligned}
\end{equation}
On the right hand side of the first row, the first factor denotes the number of single terms in the expansion of the nested commutator, the second factor shows their strength and the third corresponds to the $e^{2\lambda R}$ in the original leakage form Eq.~\eqref{eq:sm_leakage_generic_A} and we have also used $\lambda = \frac{1}{2Jkd}$ for the second row. 

One can also upper bound the leakage due to a product of nested commutators in a similar way. Such products appear in the Dyson series in Eq.~\eqref{eq:dyson_series}. Again we should focus on using Eq.~\eqref{eq:sm_leakage_generic_A} for single terms in the expansion. If we focus on a product of $j$ nested commutators with orders $n_1,n_2,\ldots,n_j$, it is easy to see that the support of single terms will be upper bounded by the sum $[(n_1+1)+(n_2+1)+\ldots+(n_j+1)]k$, this determines the new $R$ for the leakage of such term. On the other hand, the number of terms appearing will be upper bounded by the product of the number of terms in each nested commutator. Using these observations, and Eqs.~\eqref{eq:leakage_single_nested_commutator} and \eqref{eq:defintion_of_ell_n}, we deduce that the leakage from a product of nested commutators can be upper bounded by the multiplicative form 
\begin{equation}
    \ell_{n_1} \ \ell_{n_2} \ldots \ell_{n_j} \ e^{-\lambda (\Lambda' - \Lambda)}.    
\end{equation}

Using the above observation, it is now easy to see the the terms in the Dyson expansion Eq.~\eqref{eq:dyson_series} can be integrated and then recast into the exponential form.
\begin{equation}\label{eq:sm_leakage_product_final_form}
    \left\| \Pi_{>\Lambda'}  \mathcal{W}(s) \Pi_{\leq \Lambda} \right\| \leq  e^{-\lambda \left[\Lambda' - \Lambda - 2 R^{\mathcal{W}}(s) \right]},
\end{equation}
with $R^{\mathcal{W}}(s)$ defined based on the expansion in Eq.~\ref{eq:interaction_picture_E_def}:
\begin{equation}
\begin{aligned}
    \lambda R^{\mathcal{W}}(s) &= \ell_1 \frac{s^2}{2} \sum_{m,m_1} |f^{I}_{m_1,m}|\\
    &+ \ell_2 \frac{s^3}{3} \sum_{m,m_1,m_2} |f^{I}_{m_2m_1,m}|\\
    &+\ldots.
\end{aligned}    
\end{equation}
Note that we dropped $e^{-iHs}$ from the left in the expansion of $\mathcal{W}$ as it commutes with $\Pi_{>\Lambda'}$ and has norm equal to 1.

Since the expansions for $\mathcal{E}^I$ and $\mathcal{F}$ agree to the lowest nonzero order, $R^{\mathcal{W}}$ to lowest order for a Trotter-Suzuki of order $p$ can be written as:
\begin{equation}\label{eq:RI_lowest_order}
    R^{\mathcal{W}}(s) = \frac{s^{p+1}}{p+1} \, \frac{\ell_p}{\lambda} \sum_{m,\{m_i\}} |f_{\{m_i\},m}| + \mathcal{O} (s^{p+2}).
\end{equation}
$i$ in $\{m_i\}$ ranges from $1$ to $p$.
One can check that $R^{\mathcal{W}}$ scales linearly with $L$ and system size as $\ell_p$ does so.

\begin{table*}[t]
    \centering
    \begin{tabular}{|c|c|c|c|}
    \hline
         & Number of terms & Support of terms & Strength of terms \\
         \hline
       $H_m$  & $L$  & $k$ & $J$\\
         \hline
       $[H_{m_1},H_m]$  & $L k d$  & $2k$ & $J^2$\\
         \hline
       $[H_{m_2},[H_{m_1},H_m]]$  & $ Lkd (2kd)$  & $3k$ & $ J^3$ \\
       \hline
       $[H_{m_3},[H_{m_2},[H_{m_1},H_m]]]$  & $ Lkd (2kd) (3kd)$  & $4k$ & $ J^4$ \\
       \hline
    \end{tabular}
    \caption{Upper bounds for number of terms, their lengths and their strengths in nested commutators.}
    \label{tab:nested_commutator_terms}
\end{table*}

\section{Terms of nested commutators}\label{sec:nested_commutator_terms}
The number of terms each nested commutator such as $[H_{m_n}, \ldots ,[H_{m_1},H_{m}]]\ldots]$ and other properties of such terms is discussed here. 
Table \ref{tab:nested_commutator_terms} denotes upper bounds for number of terms and lengths of terms in low order nested commutators. 
Every row of the table is deduced from the previous one; e.g.~if the total number of terms in a nested commutator is $\tilde{L}_j$ and each term has a support of at most $\tilde{k}_j$, for the next level the total number of terms will be upper bounded by $\tilde{L}_j \tilde{k}_j d$ and the length of each term will be upper bound by $\tilde{k}_j+k$.
It is easy to see that for an order $n$ nested commutator, the number of terms, lengths of terms and their strengths can be upper bounded as $L (kd)^n n!$, $(n+1)k$ and $J^{n+1}$ respectively.

\section{Bounding the retained component}\label{sec:bound_retained_comp}
The error in time evolution of a low-energy state which can be quantified by the following operator norm is discussed in the main text:
\begin{equation}\label{eq:sm_error_single_step}
\begin{aligned}
    \varepsilon_\Delta(s) = \left\| \left( \mathcal{W}(s) - e^{-iHs} \right) \Pi_{\leq \Delta} \right\|.
\end{aligned}
\end{equation}
There, we broke the above norm into two separate terms as follows:
\begin{equation}\label{eq:sm_breaking_the_error}
\begin{aligned}
    \varepsilon_\Delta(s) = \left\| \Pi_{\leq \Delta'} \left( \mathcal{W}(s) - e^{-iHs} \right) \Pi_{\leq \Delta} \right\| 
    + \left\| \Pi_{> \Delta'}  \mathcal{W}(s)  \Pi_{\leq \Delta} \right\|,
\end{aligned}
\end{equation}
The second term is the leakage of the product formula that was discussed earlier.

Here, we would like to bound the following norm, which we call the retained component:
\begin{equation}\label{eq:retained_component}
\varepsilon_{<}(s)=\left\| \Pi_{\leq \Delta'} \left( \mathcal{W}(s) - e^{-iHs} \right) \Pi_{\leq \Delta} \right\| .    
\end{equation}
First, we define $\delta \mathcal{W}(s)$:
\begin{equation}\label{eq:sm_def_delta_W}
    \mathcal{W}(s)-e^{-iHs} = e^{-iHs} \delta \mathcal{W}(s),
\end{equation}
which has a derivative:
\begin{equation}
    \frac{d}{ds} \delta\mathcal{W} = \frac{d}{ds}\left(e^{iHs} \mathcal{W}(s) - 1 \right) = -i \, e^{iHs}  (\mathcal{F}(s) - H)  \mathcal{W}(s),
\end{equation}
where we have used Eq.~\eqref{eq:sm_deriv_W}. Noting the definition $\mathcal{E}=\mathcal{F}-H$, this lets us rewrite Eq.~\eqref{eq:sm_def_delta_W} as:
\begin{equation}\label{eq:integral_form_for_diff}
\begin{aligned}
    \mathcal{W}(s) - e^{-iHs} =  \int_0^s d\sigma \; e^{-iH (s-\sigma) } \, \mathcal{E}(\sigma) \, \mathcal{W}(\sigma).
\end{aligned}
\end{equation}
This equation can be used to give the total product formula error as:
\begin{equation}
    \left\|  \mathcal{W}(s) - e^{-iHs} \right\| \leq  \int_0^s d\sigma \;  \left\| \mathcal{E}(\sigma)  \right\|,
\end{equation}
where the fact that $ e^{-iH (s-\sigma) }$   and $ \mathcal{W}(\sigma)$ are unitary operators are used. This is equivalent to the bounds established in Ref.~\cite{childs2021theory}.

Getting back to evaluation of Eq.~\eqref{eq:retained_component}, we use Eq.~\eqref{eq:integral_form_for_diff} with a resolution of the identity in the form of $\Pi_{\leq \Delta'} + \Pi_{>\Delta'}$ inserted between $\mathcal{E}(\sigma)$ and $\mathcal{W}(\sigma)$:
\begin{equation}
    \varepsilon_{<}(s) = \left\| \int_0^s d\sigma \; \Pi_{\leq \Delta'} e^{-iH (s-\sigma) } \, \mathcal{E}(\sigma) \, \left( \Pi_{\leq \Delta'} + \Pi_{> \Delta'}\right) \, \mathcal{W}(\sigma) \, \Pi_{\leq \Delta} \right\|.
\end{equation}
We bound the above, by first dropping $e^{-iH (s-\sigma) }$ from the left as it commutes with the projector and has a norm equal to 1; also using the triangle inequality:
\begin{equation}\label{eq:epsilon_le_two terms}
\begin{aligned}
    \varepsilon_<(s) \leq \int_0^s d\sigma  \bigg[ \left\|\Pi_{\leq \Delta'} \, \mathcal{E}(\sigma) \,  \Pi_{>\Delta'} \, \mathcal{W}(\sigma) \, \Pi_{\leq \Delta} \right\| 
    + \left\|\Pi_{\leq \Delta'} \, \mathcal{E}(\sigma) \,  \Pi_{\leq\Delta'} \, \mathcal{W}(\sigma) \, \Pi_{\leq \Delta} 
    \right\|\bigg].
\end{aligned}
\end{equation}

We bound the two terms separately below. Beginning with the first term in Eq.~\eqref{eq:epsilon_le_two terms}, we first note:
\begin{equation}
    \int_0^s d\sigma \; \left\| \left( \Pi_{\leq \Delta'} \, \mathcal{E}(\sigma) \,  \Pi_{>\Delta'} \, \mathcal{W}(\sigma) \, \Pi_{\leq \Delta} \right)  \right\| \leq \int_0^s d\sigma \; \left\| \left( \Pi_{\leq \Delta'} \, \mathcal{E}(\sigma) \,  \Pi_{>\Delta'} \right\| \left\| \Pi_{>\Delta'} \, \mathcal{W}(\sigma) \, \Pi_{\leq \Delta} \right)  \right\|,
\end{equation}
for which the leakage bound for $\mathcal{W}$ along with the fact that $\left\|\Pi_{\leq \Delta'} \, \mathcal{E}(\sigma) \,  \Pi_{>\Delta'}\right\|$ is bounded by $\|\mathcal{E}(\sigma)\|$ can be used. For an order $p$ product formula:
\begin{equation}
\begin{aligned}
    \|\mathcal{E}(\sigma)\| \leq \sigma^p J^{p+1} L \,
    \gamma_p \sum_{m,\{m_i\}} |f_{\{m_i\},m}| +\mathcal{O}(\sigma^{p+1}),
\end{aligned}    
\end{equation}
with $\gamma_p = (kd)^p p!$ (see Table \ref{tab:nested_commutator_terms}). $i$ in $\{m_i\}$ ranges from $1$ to $p$.
As a result of this and using Eqs.~\eqref{eq:sm_leakage_product_final_form} and \eqref{eq:RI_lowest_order}, the whole first term can be bounded as:
\begin{equation}\label{eq:leakage_retained_term_leakout}
\begin{aligned}
    \int_0^s d\sigma \; \left\| \left( \Pi_{\leq \Delta'} \, \mathcal{E}(\sigma) \,  \Pi_{>\Delta'} \, \mathcal{W}(\sigma) \, \Pi_{\leq \Delta} \right)  \right\| \leq 
    \xi_1 \, e^{-\lambda(\Delta' - \Delta - R_1)},
\end{aligned}
\end{equation}
where $\xi_1 = \mathcal{O}(s^{p+1} J^{p+1} L)$ and $R_1 = \mathcal{O}(s^{p+1}J^{p+2} L)$. The integral over $\sigma$ has been upper bounded by multiplying the maximum value of the integrand (which happens at $\sigma=s$) by $s$. $\xi_1$ can be absorbed into $R_1$ resulting in a logarithmic correction in $s$ and $L$ only. This means that, as we will discuss more below, this leakage term can become subdominant with respect to the first term by taking $\Delta'-\Delta = \mathcal{O}(s^{p+1} L)$ up to polylogarithmic factors in $s$ and $L$. Note that the same condition also guarantees suppression of the second term in Eq.~\eqref{eq:sm_breaking_the_error}, i.e.~the leakage of the product formula according to Eqs.~\eqref{eq:sm_leakage_product_final_form} and \eqref{eq:RI_lowest_order}.

Turning to the second term in Eq.~\eqref{eq:epsilon_le_two terms}, this time we bound $\|\Pi_{\leq\Delta'} \, \mathcal{W}(\sigma) \, \Pi_{\leq \Delta}\|$ by the norm of $\mathcal{W}(\sigma)$ which is equal to 1. Thus the second row of Eq.~\eqref{eq:epsilon_le_two terms} can be upper bounded by:
\begin{equation}\label{eq:integral_of_E_for_low_energy}
    \int_0^s d\sigma \; \left\| \Pi_{\leq \Delta'} \, \mathcal{E}(\sigma) \,  \Pi_{\leq\Delta'}  \right\|
\end{equation}
First we discuss this general form. It is showing us that up to an error $\vartheta$, the expression in Eq.~\eqref{eq:integral_of_E_for_low_energy} can represent $\varepsilon_\Delta(s)$ in Eq.~\eqref{eq:sm_error_single_step} if $\Delta'$ is chosen such that $\Delta'-\Delta = \mathcal{O}(s^{p+1} L + \log(1/\vartheta))$ up to polylogarithmic factors in $s$ and $L$. 
The condition on $\Delta'$ ensures that both of the leakage terms, i.e. Eq.~\eqref{eq:leakage_retained_term_leakout} and the second term on the right hand side of \eqref{eq:sm_breaking_the_error}, are subdominant error terms compared with Eq.~\eqref{eq:integral_of_E_for_low_energy} and in fact bounded by $\vartheta$.

Note that this is a general result and it is not dependent on the terms of the Hamiltonian being positive semidefinite. It is showing that for a product formula the error for low energies can be computed by projecting the operator $\mathcal{E}$ from both sides to a low-energy subspace, provided that $\Delta'$ is taken to be large enough as above. If this equation is considered to lowest order in $s$, terms with nested commutators of order $p$ such as $[H_{m_p}, \ldots ,[H_{m_1},H_m]]\ldots]$ remain, and we will obtain:
\begin{equation}\label{eq:error_nested_commutator_projected}
    \frac{s^{p+1}}{p+1} \, \sum_{m=1}^M \sum_{\{m_i\} } f_{\{m_i\},m} \  \left\| \Pi_{\leq \Delta'} \ [H_{m_p}, \ldots ,[H_{m_1},H_m]]\ldots] \  \Pi_{\leq\Delta'}  \right\| + \mathcal{O}(s^{p+2}).
\end{equation}
Compare this result with Eq.~\eqref{eq:sm_error_operator_norm}. In the main text the $f$ coefficients are dropped as they are $\mathcal{O}(1)$.

The form in Eq.~\eqref{eq:error_nested_commutator_projected} can also be recast in a way that is useful for Hamiltonian with terms $H_m$ being positive semidefinite.
From here on, a number of energy cutoffs denoted as $\Delta'<\Delta_1 < \Delta_2 < \ldots < \Delta_{p-1} = \Delta_f$ should be introduced in order to use leakage bounds. Using resolutions of identity we can recursively reduce the nested commutator by noting that:
\begin{equation}\label{eq:bounding_retained_nested}
\begin{aligned}
    \left\| \Pi_{\leq \Delta'} \, [H_{m_p}, \ldots ,[H_{m_1},H_m]]\ldots] \,  \Pi_{\leq\Delta'}  \right\| &\leq \left\| \Pi_{\leq \Delta'} \, H_{m_p} (\Pi_{\leq \Delta_1} + \Pi_{> \Delta_1}) [H_{m_{p-1}}, \ldots ,[H_{m_1},H_m]]\ldots] \,  \Pi_{\leq\Delta'}  \right\| \\
    &+ \left\| \Pi_{\leq \Delta'} \, [H_{m_{p-1}}, \ldots ,[H_{m_1},H_m]]\ldots]    (\Pi_{\leq \Delta_1} + \Pi_{> \Delta_1}) H_{m_p}\,  \Pi_{\leq\Delta'}  \right\| \\
    &\leq 2  \left\| \Pi_{\leq \Delta_1} \, H_{m_p} \Pi_{\leq \Delta_1} \right\|   \left\|  \Pi_{\leq \Delta_1} \, [H_{m_{p-1}}, \ldots ,[H_{m_1},H_m]]\ldots]    \Pi_{\leq \Delta_1} \right\| \\
    &+ 2  \left\| \Pi_{> \Delta_1} \, H_{m_p} \Pi_{\leq \Delta'} \right\|   \left\|  \Pi_{> \Delta_1} \, [H_{m_{p-1}}, \ldots ,[H_{m_1},H_m]]\ldots]    \Pi_{\leq \Delta'} \right\| 
\end{aligned} 
\end{equation}
This procedure can be continued until all nested commutators are resolved; the terms on the row before the last are projected to a low-energy subspace, but we also need to bound the terms on the last row of Eq.~\eqref{eq:bounding_retained_nested}, which are leakages; we can use Eq.~\eqref{eq:leakage_single_nested_commutator} to this end.
\begin{equation}
    \left\| \Pi_{> \Delta_1} \, H_{m_p} \Pi_{\leq \Delta'} \right\|   \left\|  \Pi_{> \Delta_1} \, [H_{m_{p-1}}, \ldots ,[H_{m_1},H_m]]\ldots]    \Pi_{\leq \Delta'} \right\| \leq \ell_0 \ell_{p-1} e^{-2\lambda(\Delta_1 - \Delta')},
\end{equation}
with $\ell_n$ defined in Eq.~\eqref{eq:defintion_of_ell_n}. Note that the prefactor $\ell_0 \ell_{p-1}$ scales as $\mathcal{O}(L^2)$.

This procedure should be continued until only norms of single terms of the Hamiltonian (without commutators) projected to low energies remain: 
\begin{equation}\label{eq:single_nested_coomutator_retained_bounded}
\begin{aligned}
    \frac{s^{p+1}}{p+1} \bigg( &2 \ell_0 \ell_{p-1} \ e^{-2\lambda(\Delta_1 - \Delta')} \\
    &+2^2 \left\| \Pi_{\leq \Delta_1} \, H_{m_p} \Pi_{\leq \Delta_1} \right\| \ \ell_0 \ell_{p-2} \ e^{-2\lambda(\Delta_2 - \Delta_1)} \\
    &+ \ldots\\
    &+2^{p-1} \left\| \Pi_{\leq \Delta_1} \, H_{m_p} \Pi_{\leq \Delta_1} \right\| \ldots \left\| \Pi_{\leq \Delta_{p-2}} \, H_{m_1} \Pi_{\leq \Delta_{p-2}} \right\| \ \ell_0 \ell_{1} \ e^{-2\lambda(\Delta_{p-1} - \Delta_{p-2})}\\
    &+ 2^p  \left\| \Pi_{\leq \Delta_1} \, H_{m_p} \Pi_{\leq \Delta_1} \right\| \ldots \left\| \Pi_{\leq \Delta_{p-2}} \, H_{m_1} \Pi_{\leq \Delta_{p-2}} \right\| \left\| \Pi_{\leq \Delta_f} \, H_{m} \Pi_{\leq \Delta_f} \right\|   \bigg).
\end{aligned}
\end{equation}
Ultimately we would be interested in choosing the energy cutoffs in a way that the exponential factors make all the terms subleading with respect to the last term in Eq.~\eqref{eq:single_nested_coomutator_retained_bounded}. This can be done by taking the differences between adjacent cutoffs to be large enough, since all the prefactors of the exponentials scale as powers of $L$, the difference between adjacent cutoffs suffices to grow only logarithmically with $L$. This means that $\Delta_f$ and $\Delta'$ only differ by polylogarithmic factors and thus $\Delta_f-\Delta= \mathcal{O}(s^{p+1} L)$ up to polylogarithmic factors in $s$ and $L$ will guarantee suppression of all the leakage terms. Note that it is simple to extend this analysis to include orders beyond the leading order by adding more energy cutoffs, and $\Delta_f$ will only change polylogarithmically.

This leaves us with the following two forms for the error:
\begin{itemize}
    \item Positive semidefinite $H_m$ terms: each operator norm in the above relation can be bounded by $\Delta_f$ for Hamiltonians with positive semidefinite terms $H_m$. As a result, the error in a single step takes the form:
    \begin{equation}\label{eq:sm_error_single_step_pos}
        \varepsilon_\Delta(s) = \mathcal{O}(s^{p+1} \Delta_f^{p+1}).
    \end{equation}
    Notice that while we can add constants to the $H_m$'s to make them positive semidefinite, that would likely make $\Delta$ acquire an undesirable scaling with $N$ (this is similar to the argument in \cite{csahinouglu2021hamiltonian}), cancelling the advantage of the low-energy error analysis.

    \item Non-positive semidefinite $H_m$ terms: in this case, we need to define the quantity  $\tilde{\Delta}_f = \max_m \min_C \| \Pi_{\leq \Delta_f} (H_m + C) \Pi_{\leq \Delta_f} \|$ (this is similar to the discussion in \cite{csahinouglu2021hamiltonian}) with $C$ a constant. Note that we have added a constant, because the above bound is found through the use of commutators and adding constants leaves them unchanged. In this case, the error in a single step will take the form:
    \begin{equation}\label{eq:sm_error_single_step_nonpos}
        \varepsilon_\Delta(s) = \mathcal{O}(s^{p+1} \tilde{\Delta}_f^{p+1}).
    \end{equation}
\end{itemize}
The error in Eq.~\eqref{eq:sm_error_single_step_pos} has been used in the main text to calculate the cost of performing low-energy time evolution for relevant systems. However, the cost scaling that corresponds to the error in Eq.~\eqref{eq:sm_error_single_step_nonpos} is not as simple to calculate. The main problem is that in this case since $H_m$ terms are not taken to be positive-semidefinite, it is not simple to study the behavior of $\tilde{\Delta}_f$ with system properties such as its size. Characterization of this parameter for Hamiltonians of interest can be an compelling subject of future study.

\end{document}